\documentstyle[12pt,epsfig]{article}

 1
\def\as{\alpha_{\rm S}}

\def\citenum#1{{\def\@cite##1##2{##1}\cite{#1}}}
\def\citea#1{\@cite{#1}{}}
\def\as{\alpha_{\rm S}}

\def\g{\gamma}

\def\O{\Omega}

\def\s{\sigma}

\def\({\left(}
\def\){\right)}

\def\citenum#1{{\def\@cite##1##2{##1}\cite{#1}}}
\def\citea#1{\@cite{#1}{}}

\def\l1vt{\vec{l_{1\perp}}}

\def\rt{r_{\perp}}
\def\bt{b_{\perp}}
\def\rt2{$r^2_{\perp}$}
\def\bt2{$b^2_t$}

\def\jol1{$J_0(\,l_{1\perp}\,r_{\perp}\,)$}

\def\citea#1{\@cite{#1}{}}

\def\beq{\begin{equation}}
\def\eeq{\end{equation}}
\def\bea{\begin{eqnarray}}
\def\eea{\end{eqnarray}}

\def\eq#1{{Eq.~(\ref{#1})}}

\def\bbbz{{\mathchoice {\hbox{$\sf\textstyle Z\kern-0.4em Z$}}
{\hbox{$\sf\textstyle Z\kern-0.4em Z$}}
{\hbox{$\sf\scriptstyle Z\kern-0.3em Z$}}
{\hbox{$\sf\scriptscriptstyle Z\kern-0.2em Z$}}}}
\def\npb#1#2#3{    {\it Nucl. Phys. }{\bf B#1} (19#2) #3}
\def\plb#1#2#3{    {\it Phys. Lett. }{\bf B#1} (19#2) #3}
\def\prd#1#2#3{    {\it Phys. Rev. }{\bf D#1} (19#2) #3}

\def\prl#1#2#3{    {\it Phys. Rev. Lett. }{\bf #1} (19#2) #3}

\def\zpc#1#2#3{    {\it Z. Phys. }{\bf C#1} (19#2) #3}

\def\sjnp#1#2#3{   {\it Sov. J. Nucl. Phys. }{\bf #1} (19#2) #3}

\def\l{\lambda}

\begin{document}
\begin{titlepage}
\noindent
\begin{flushright}
ANL - HEP - PR - 96 -52\\
hep - ph 9607\\
 1 July 1996\\
\end{flushright}
\begin{center}

{\Large\bf{FROISSART BOUNDARY}}\\[1.4ex]
{\Large \bf { FOR DEEP INELASTIC STRUCTURE}}\\[1.4ex]
{\Large \bf { FUNCTIONS}}\\[9ex]
{\large \bf { A. L.
Ayala  F$^{\underline{o}}$ ${}^{a)\,b)}$${}^*$\footnotetext{ ${}^*$ E-mail:
ayala@if.ufrgs.br}
,
 M. B. Gay  Ducati ${}^{a)}$$^{**}$\footnotetext{${}^{**}$
E-mail:gay@if.ufrgs.br}}}\\
 and\\
{ \large \bf{ E. M. Levin ${}^{c)\,d)}{}^{\dagger}$
\footnotetext{${}^{\dagger}$ E-mail:
levin@fnal.gov;leving@ccsg.tau.ac.il} }}\\[1.5ex]

{\it ${}^{a)}$Instituto de F\'{\i}sica, Univ. Federal do Rio Grande do Sul}\\
{\it Caixa Postal 15051, 91501-970 Porto Alegre, RS, BRAZIL}\\[1.5ex]
{\it ${}^{b)}$Instituto de F\'{\i}sica e Matem\'atica, Univ. 
Federal de Pelotas}\\
{\it Campus Universit\'ario, Caixa Postal 354, 96010-900, Pelotas, RS,
BRAZIL}\\
{\it ${}^{c)}$ Physics Division, Argonne National Laboratory }
\\
{\it Argonne, IL 60439, USA}\\
{\it$ {}^{d)}$ Theory Department, Petersburg Nuclear Physics Institute}\\
{\it 188350, Gatchina, St. Petersburg, RUSSIA}\\[6.5ex]
\end{center}
{\large \bf Abstract:}
In this letter we derive the Froissart boundary in QCD  for the deep inelastic 
structure function in low $x$ kinematic region. We show that
 the comparison of the 
Froissart boundary with the new HERA experimental data  gives rise to a challenge 
for QCD to explain the matching between the deep inelastic scattering 
and real photoproduction process.
\end{titlepage}

\section{Introduction.}
The new HERA data \cite{HERASIGMA} show the steep $x$-dependence of the 
total cross section in the deep inelastic scattering ( DIS ) 
 of virtual photon off 
a proton ( $ \s_{tot} ( \g^* p ) $ ). Approximately, $\s( \g^{*} p )\,\propto 
\, x^{- 0.2} $ at small $x$ ( $ 10^{-2}\,\leq\,x\,\leq\,10^{-5}$ ). 
Surprisingly, this energy rise holds at rather small photon virtualities 
 ( $Q^2 \,\approx\, 1 - 2 GeV^2 $ ). At first sight it 
 means that in HERA kinematic 
region we still have sufficiently diluted parton cascade and the 
parton-parton interaction which shall stop the increase of the parton 
density \cite{GLR}  is still rather small.    On the other hand, the 
probability of the parton - parton interaction \cite{GLR}
($\kappa$ )
%\footnote{In Ref.\cite{GLR}  the notation for $\kappa$ was W, 
%but here we  use $\kappa$ to avoid a misunderstanding since W is the 
%standard notation for the energy of interaction in DIS.} 
is equal to
\beq 
\label{I1}
\kappa\,=\,x G ( x,Q^2 ) \frac{\s(GG)}{\pi R^2}\,=\,\frac{N_c \as \pi}{2 
Q^2 R^2}\,x G ( x,Q^2 )\,\,,
\eeq
\begin{figure}[b]
\centerline{\epsfig{file=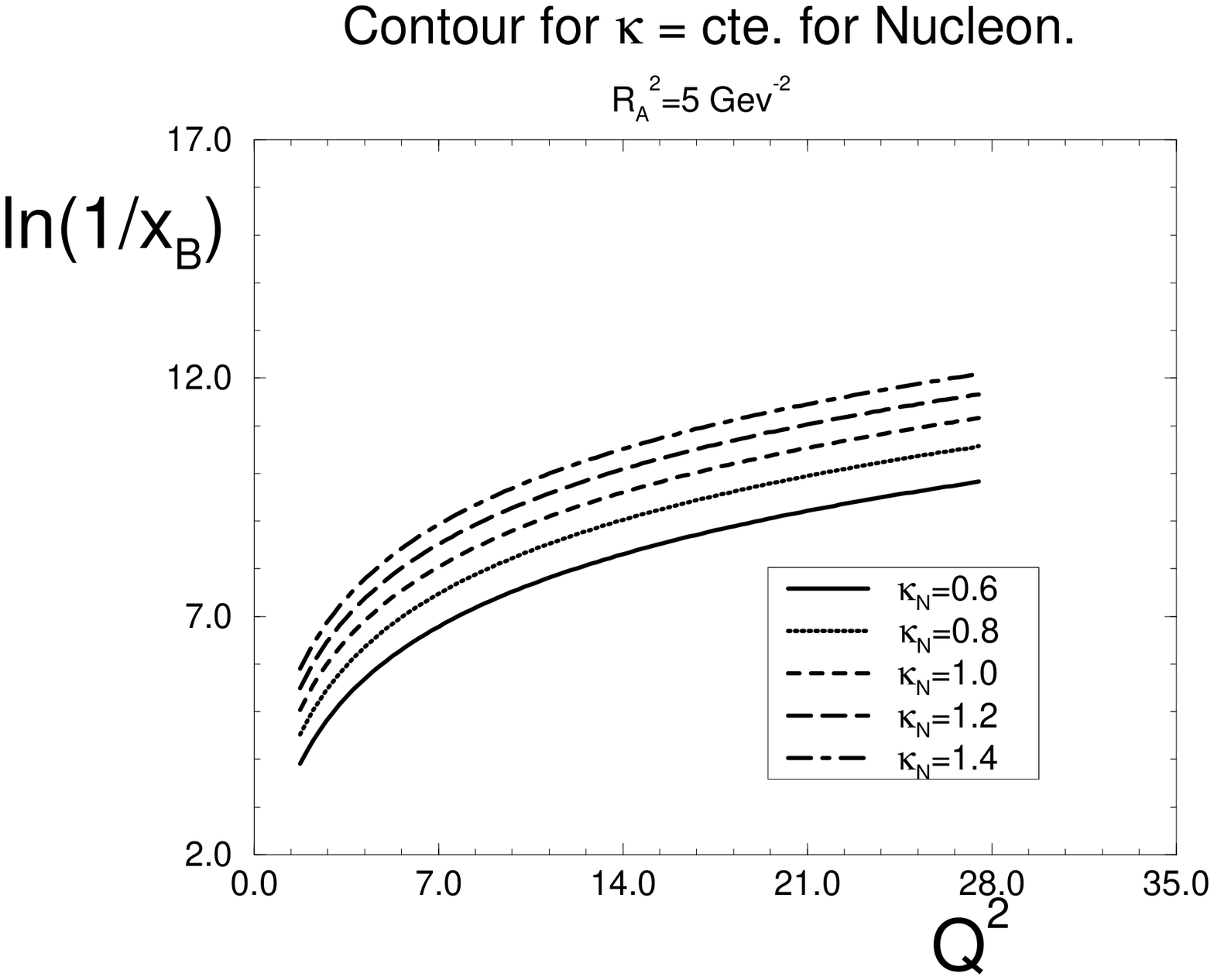,width=90mm}} 
\caption{Contour plot for $\kappa$ for $R^2= \, 5 Gev^{-2}$.}
\label{Fig.1}
\end{figure}
where $x G(x,Q^2)$ is the number of partons ( gluons) in the parton 
cascade, $\s$ is the cross section of parton-parton interaction and 
$R^2$ is the size of a hadron. The numerical factor in \eq{I1} has been
 evaluated by Mueller and Qiu \cite{MUQI} and has been confirmed in many 
 further publications \cite{KAPPA}. Fig.\ref{Fig.1} shows the contour plot for 
$\kappa$ using the GRV parameterization \cite{GRV} for the gluon 
structure function and the value of $R^2 = 5 GeV^{-2}$. We will argue 
a bit later that this value of $R^2$ follows directly from HERA 
measurement of the diffraction production of J/$\Psi$ meson 
\cite{HERAPSI}. One can see that $\kappa$ reaches $\kappa$ = 1 at HERA 
kinematic region, meaning shadowing corrections take place. 
Therefore, the situation looks very controversial.

The goal of this letter is to derive the Froissart boundary for the 
deep inelastic structure function. This should clarify when the 
shadowing corrections to the deep inelastic 
process become important. Some attempts have been made to derive 
the geometrical limit ( 
the Froissart boundary) during the last three decades ( see 
Refs.\cite{GRIB} \cite{BJ} and lectures \cite{AFS} for update review on 
the subject) assuming that a target is black for the dominant hadronic
  component in the wave function of the virtual photon. We derive the 
Froissart boundary for the deep inelastic  processes  assuming 
the GLAP evolution equation for the parton densities and the colour 
dipole picture of interaction proposed by A. Mueller \cite{MU90} 
\cite{MUDI} ( see also Refs.\cite{LR87} \cite{LR90} \cite{NZ} where many 
of aspects of the Mueller approach have been foreseen).

\section{Unitarity constraint for DIS.}

\subsection{$s$ - channel unitarity ( general formulae).}

\begin{figure}
\centerline{\epsfig{file=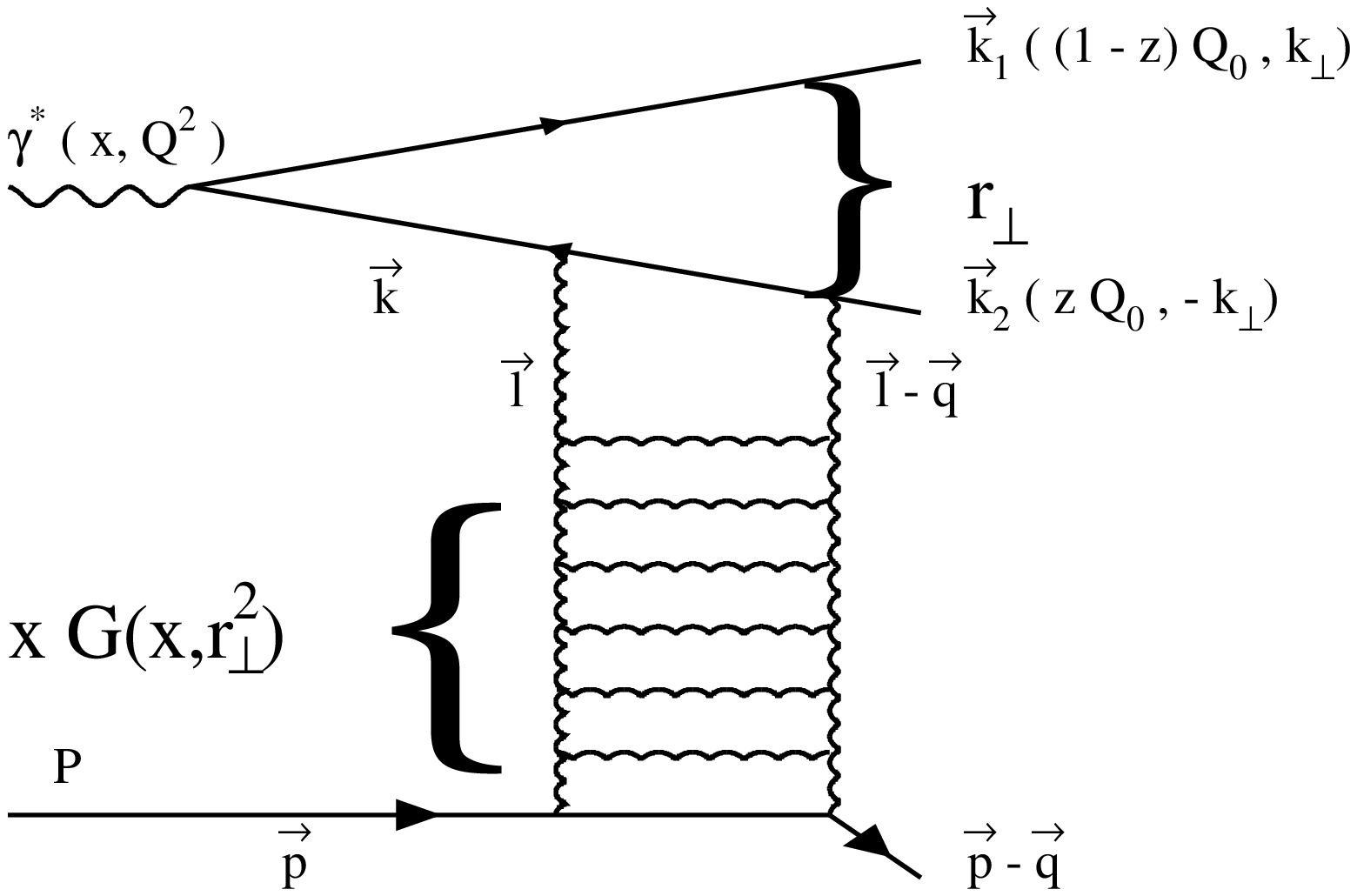,width=90mm}} 
\caption{DIS in the target rest frame.}
\label{Fig.2}
\end{figure}

The unitarity constraint can easily be derived considering the DIS 
in the frame where a target is at rest. In this frame the virtual photon
at high energy ( small $x$ )
decays in quark - antiquark ( $\bar q q $ ) 
 pair  long before the interaction with the 
target. The $\bar q q $ system
 traverses the target with  fixed transverse distance $r_{\perp}$ 
between quark and antiquark \cite{LR87} \cite{MU90}. Indeed, $r_{\perp}$ 
can vary by amount $\Delta r_{\perp}\,\propto\,R \,\frac{k_{\perp}}{Q_0}$,
where $Q_0$ denotes the energy of the $\bar q q$ pair or the 
virtual photon in the target rest frame, $R$ is the size of the 
target, and the quark momentum is $k_{\perp}\,\propto 
\,\frac{1}{r_{\perp}}$  ( see Fig.\ref{Fig.2} ). Therefore
\beq \label{11}
\Delta 
r_{\perp}\,\,\propto\,\,R\,\frac{k_{\perp}}{Q_0}\,\,\ll\,\,r_{\perp}\,\,,
\eeq
which in terms of $x$ has the following form:
\beq \label{12}
x\,\,\ll\,\,\frac{1}{ 2 m R}\,\,.
\eeq   
The cross section for the DIS can be written in the form:
\beq \label{13}
\s(\g^* p )\,\,=\,\,\int^1_0\,d z \,\int\,d^2 r_{\perp} \,\Psi ( z, 
r_{\perp} ) \,\s_{tot}( zQ_0, r^2_{\perp} )\,\Psi^*(z, r_{\perp})\,\,,
\eeq
cross section for $\bar q q $ interaction with the 
target, $z$ is the fraction of energy of the photon ( $Q_0$ ) carried by 
quark and $r_{\perp}$ is the transverse separation between quark and 
antiquark. $\Psi$ is the wave function of $\bar q q $ - pair in the 
virtual photon. This wave function is well known \cite{MU90} 
\cite{NZ}  and $|\Psi_T |^2$ for transverse polarized photon and for 
massless quarks  is equal to
$$ 
| \Psi_T |^2\,\,=\,\,\frac{\as^{em} N_c}{2 \pi^2} 
\,\sum^{N_f}_{1}\,Z^2_f 
\,[\, z^2 + ( 1 - z )^2\,]\,{\bar Q}^2\, K^2_1 (\,\bar 
Q\,r_{\perp}\,)\,\,,
$$
where $K_1$ is the modified Bessel function, ${\bar Q}^2\,=\,z(1-z)Q^2$ ,
 $N_f$ is the number of massless quarks and $Z_f$ is the fraction of the 
charge carried by the quark.

The main contribution in  \eq{13} (see Ref.\cite{MU90} for details ),
 which corresponds to the GLAP 
evolution, comes  from the region $ \bar Q \,r_{\perp} \,\leq\,1 $ and
$z(1 - z) \,\leq\,\frac{1}{Q^2\,r^2_{\perp}}\,\leq\,\frac{1}{4}$. 
In this case the integral over $z$ can be taken explicitly. Since 
$z\,\,\ll\,\,1$,  it can be reduced to the integral
\beq \label{INT}
\int^1_0 d z {\bar Q}^2 \,K^2_1 ( \bar Q \,r_{\perp} )\,\,\rightarrow
\,\,2 \,Q^2 \int^{\infty}_0 \,z\,d z \,K^2_1 (\sqrt{ z}\,Q\,r_{\perp} )\,=\,
\frac{8}{3\,Q^2\,r^4_{\perp}}\,\,,
\eeq
with $Q^2 \,r^2_{\perp}\,\geq\,4 $.
Finally, we have
\beq \label{14}
\s(\g^* p )\,\,=\,\,\frac{4\, N_c\, \as^{em}}{3\,\pi} \,\sum^{N_f}_{1}\,Z^2_f
\,\, \frac{1}{Q^2}\,\int^{\infty}_{\frac{4}{Q^2}}\,\,\frac{d 
r^2_{\perp}}{r^4_{\perp}}\,\s_{tot}( 
\frac{Q_0}{Q^2\,r^2_{\perp}},r^2_{\perp})\,\,.
\eeq
At high energies ( low $x$ ) we can restrict ourselves by summing only 
$( \as \ln (1/x) )^n $ contribution in each $\as^n$ order of 
perturbative QCD ( so called leading log(1/$x$) approximation (LL($x$)A)).
In the framework of the  LL(x)A we  can safely replace the argument of $\s$ 
in 
\eq{14} by $x$. Taking into account the relation between the cross 
section and $F_2$ structure function, namely 
$\s_{tot}(\g^* p)\,=\,\frac{4 \pi^2 \as}{Q^2} F_2 ( x, Q^2)$,
 the final formula has a form \cite{MU90}:
\beq \label{15}
F_2(x,Q^2)\,\,=\,\,\frac{ N_c }{12 \pi^3} \,\sum^{N_f}_{1}\,Z^2_f
\,\,\int^{\infty}_{\frac{1}{Q^2}}\,\,\frac{d r^2_{\perp}}{r^4_{\perp}}\,
\s_{tot}(x ,\frac{r^2_{\perp}}{4})\,\,.
\eeq
The  total cross section for $\bar q q $ scattering can be written as
\beq \label{16}
\s_{tot}(x ,\frac{r^2_{\perp}}{4})\,\,=\,\,2\,\int\,d^2 
b_{\perp}\,Im\,a(x,r_{\perp}, b_{\perp})\,\,,
\eeq
where $a$ is the amplitude for elastic scattering of $\bar q q $ in 
impact parameter ($b_{\perp}$) space which is defined as 
\beq \label{17}
a(x,r_{\perp}, b_{\perp})\,\,=\,\,\frac{1}{ 2 \pi} \,\int\,d^2 
q_{\perp} 
\,e^{-\,i\,{\vec{q}}_{\perp} \cdot {\vec{b}}_{\perp}}\,f(x, r_{\perp}, t = - 
q^2_{\perp})\,\,,
\eeq
where ${\vec{q}}_{\perp}$ is the momentum transfer (see Fig.\ref{Fig.2}).
In this representation
\beq \label{18}
\s_{el}\,\,=\,\,\int\,d^2 b_{\perp}\,| a(x,r_{\perp}, b_{\perp}) |^2\,\,.
\eeq
The amplitude is normalized such that:
\beq \label{19}
\frac{ d \s}{d t}\,\,=\,\,\pi \,| f(x, r_{\perp}, t = - q^2_{\perp}) 
|^2\,\,;
\eeq
\beq \label{120}
\s_{tot}\,\,=\,\,4 \pi Im f(x, r_{\perp}, t = 0 )\,\,.
\eeq  

The $s$ - channel unitarity establishes the relationship between the 
elastic amplitude ($a$) and the contribution of all inelastic process
( $G_{in}(x, r_{\perp},b_{\perp})$ ) and has the form:
\beq \label{121}
2\,Im \,a(x, r_{\perp}, b_{\perp}) \,\,=\,\,| a(x, r_{\perp}, b_{\perp}) |^2
\,\,+\,\, G_{in}(x, r_{\perp},b_{\perp})\,\,.
\eeq
The solution of the unitarity constraint of \eq{121} is very simple if 
we assume that the elastic amplitude is predominantly imaginary at high 
energy. Indeed, one can check that the general solution of \eq{121}
in this case has  a form:
\beq \label{122}
a\,=\,i\,\{\,1 \,\,-\,\,e^{ - \,\frac{1}{2}\,
\O( x,r_{\perp}, b_{\perp} )}\,\}\,\,;
\eeq
\beq \label{123}
G_{in}\,\,=\,\,\{\,1 \,\,-\,\,e^{ - \,
\O( x,r_{\perp}, b_{\perp} )}\,\}\,\,.
\eeq
where $\O( x,r_{\perp}, b_{\perp} )$ is the opacity function. Substituting \eq{122} in \eq{16} and \eq{15}, we obtain
\beq \label{F2}
F_2(x,Q^2)\,\,=\,\,\frac{ N_c}{6\,\pi^3}\,\sum^{N_f}_{1}\,\,Z^2_f\,\, 
 \,\int^{\infty}_{\frac{1}{Q^2}} \,\frac{ d r^2_{\perp}}{r^4_{\perp}}
\, \int\,d^2 b_{\perp}\,\,\{\,
1\,\,-\,\,e^{-\,\frac{1}{2} \,\O ( x, r_{\perp}, b_{\perp} )}\,\}\,\,.
\eeq 

\subsection{ Properties  of $\O$.}
The opacity $\O$ is an arbitrary real function  which requires
 more detailed QCD calculations in order to be found ( see for example Refs. 
\cite{GLR} \cite{MUQI} 
\cite{LR87}
\cite{MU90})
and/or use the general property of 
analyticity  and crossing symmetry ( see Refs. \cite{FRST} \cite{MAR} ).

Let us recall what is known about $\O$:

1. If $\O\,\ll\,1$ one can expand the exponent in \eq{122} and \eq{15}
can be reduced to a simple form:
\beq \label{124}
F_2(x,Q^2)\,\,=\,\,\frac{ N_c }{12\, \pi^3} \,\sum^{N_f}_{1}\,Z^2_f
\,\,\int_{\frac{1}{Q^2}}^{\infty} \frac{d 
r^2_{\perp}}{r^4_{\perp}}\,\int \,d^2 b_{\perp} 
\O(x ,\frac{r^2_{\perp}}{4}, b_{\perp})\,\,.
\eeq
Differentiating over $\ln Q^2$ and comparing \eq{124} with the GLAP 
evolution equations in the region of small $x$ one can obtain for 
$\O$ the following result ( see Refs. \cite{LR87} \cite{MU90} \cite{KOP}
\cite{BFS} \cite{GLMPSI}  for details):
\beq  \label{O}
\int\, d^2 b_{\perp} \O (x,\frac{1}{ Q^2}, b_{\perp}) \,\,=\,\,
\frac{4\,\pi^2\,\as(Q^2)}{3 \,Q^2}\,\,x \,G(x,Q^2)\,\,,
\eeq
where $x G(x, Q^2) $ is the gluon structure function of the proton.

\begin{figure}
\begin{tabular}{c c}
\epsfig{file=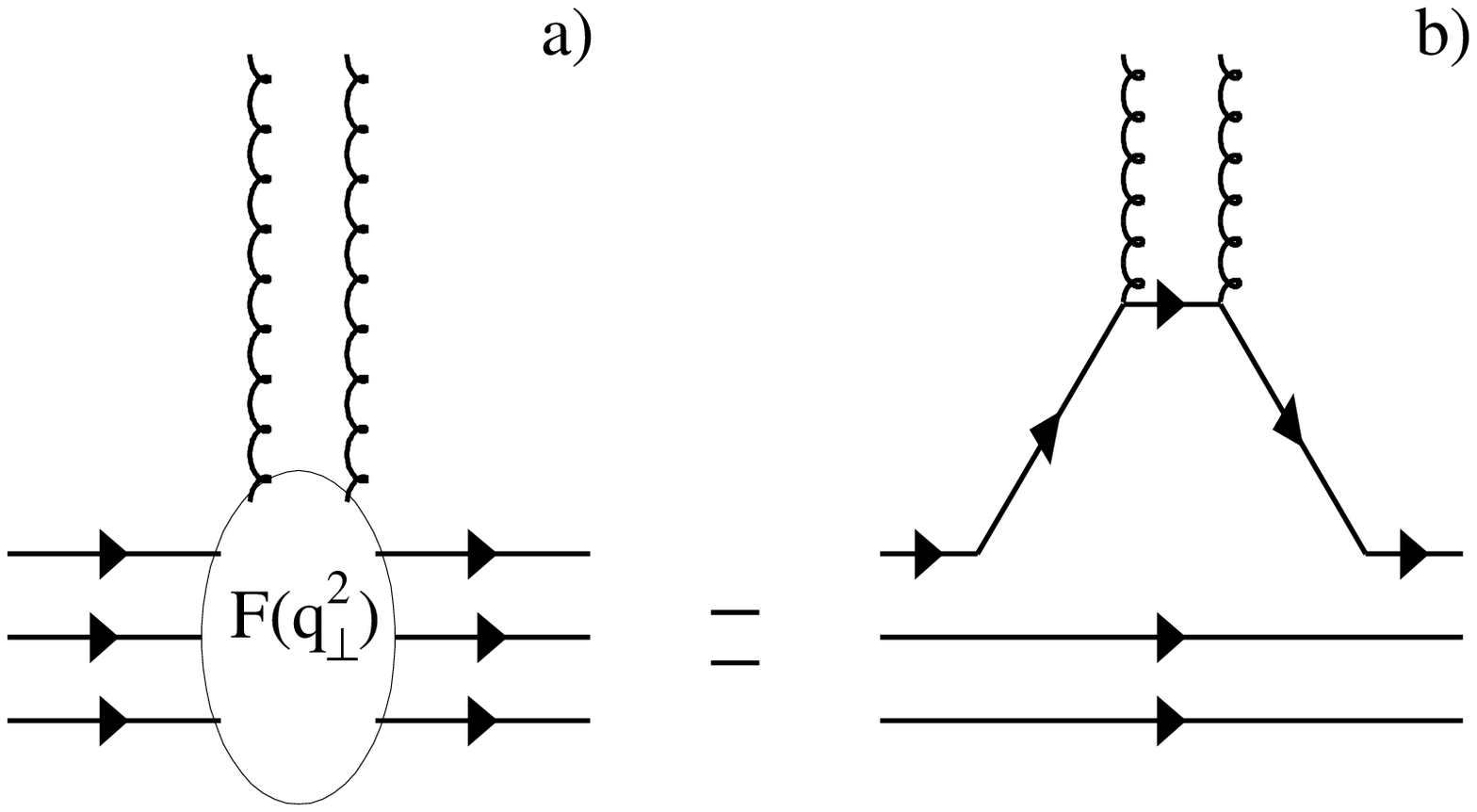,width=90mm}&\epsfig{file=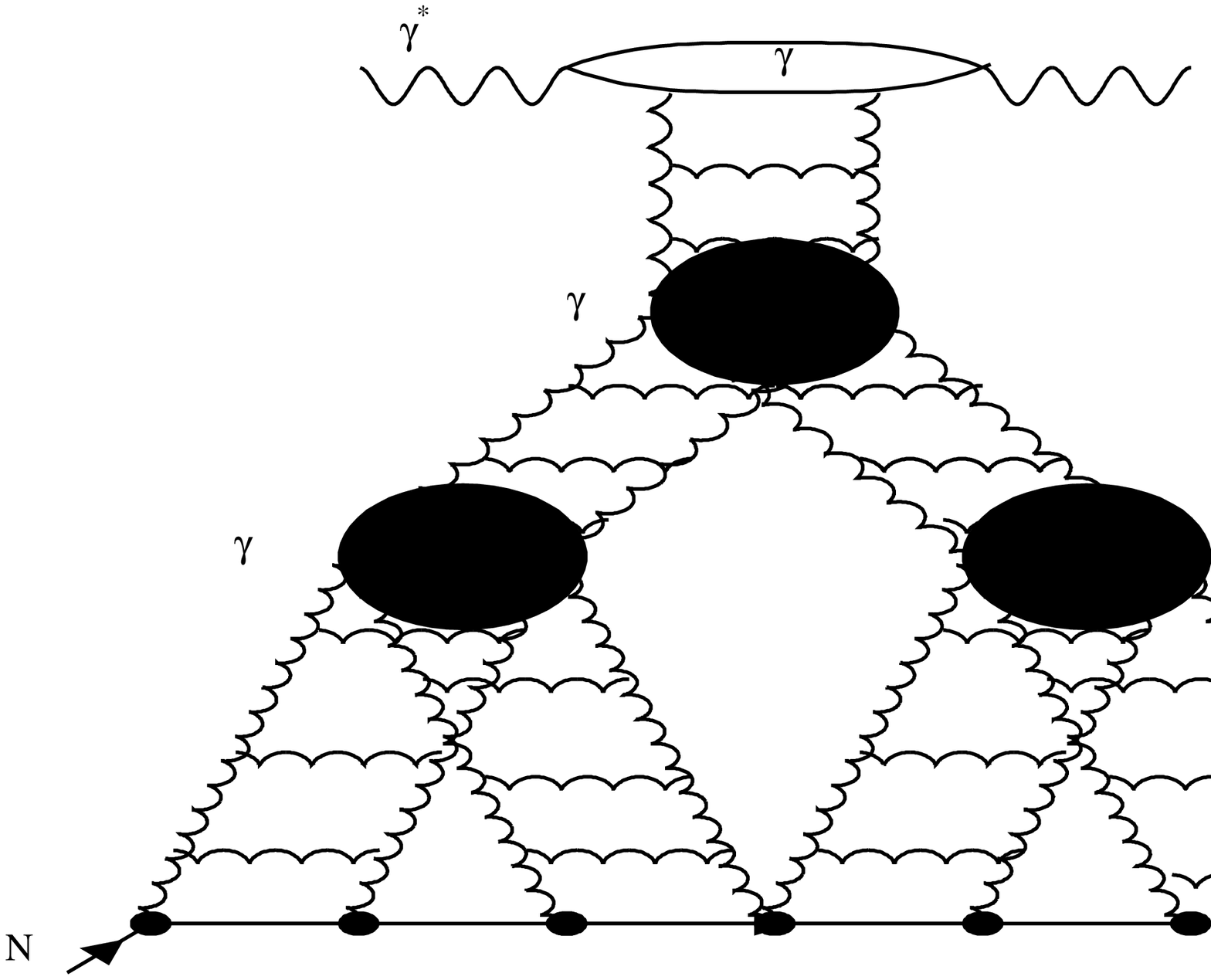,width=60mm}\\
\end{tabular} 
\caption{Two gluon proton form factor in the additive quark model (a) 
and b));
 c) The ``fan"
diagrams for DIS.}
\label{Fig.3}
\end{figure}

2. In the GLAP evolution equations the $b_{\perp}$-dependence of
the  deep inelastic structure function can be factorized ( see Refs. 
\cite{GLR} \cite{LR90})  in the form:
\beq \label{125}
\O\,\,=\,\,\frac{4\,\pi^2\,\as(Q^2)}{3 \,Q^2}\,\,x 
\,G(x,Q^2)\,S(b^2_{\perp})\,\,,
\eeq
with the profile function $S(b^2_{\perp})$ equal
\beq \label{126}
S(b^2_{\perp})\,\,=\,\,\frac{1}{(2 \pi)^2}\,\int\,d^2 
q_{\perp} \,e^{-\,i\,{\vec{q}}_{\perp} \cdot {\vec{b}}_{\perp}}\,
F(q^2_{\perp})\,\,,
\eeq
where $F(t)$ is the two gluon
 form factor of the proton pictured in  Fig.\ref{Fig.3}a.
Using, for example, the additive quark model (AQM) we can expect that this 
form factor is equal to the electromagnetic proton form factor ( see 
Fig.\ref{Fig.3}b).  Taking two different form of the proton form factor: the 
dipole ( $F_{dip}\,=\,\frac{1}{(\,1\,+\,R^2 q^2_{\perp}/8\,)^2}$ ) and 
exponential ( $ F_{exp}\,=\,exp(- \frac{1}{4} R^2 q^2_{\perp}  $ )
 ones, one can find
 two different profile functions, namely:
\beq \label{127}
 S_{dip}( b_{\perp} ) \,\,=\,\,\frac{2}{\pi\,R^2 }\,2 \sqrt{2}
\frac{b_{\perp}}{R}\,K_1( 2 \sqrt{2} \frac{b_{\perp}}{R} )\,\,; 
\eeq
and
\beq \label{128}
S_{exp} ( b_{\perp} )\,\,=\,\,\frac{1}{\pi\,R^2}\,e^{ 
-\,\frac{b^2_{\perp}}{R^2}}\,\,;
\eeq
with normalization $ \int d^2 b_{\perp} S(b_{\perp}) \,\,=\,\,1 $.

3.  We can recover the eikonal (Glauber) model for the shadowing 
corrections (SC ) if we postulate \eq{125} for $\O$ with the profile 
function $S(b_{\perp})$ from \eq{127} or \eq{128} for any values of 
$b_{\perp}$. The physical meaning of this assumption is very simple: the final 
inelastic state is an uniform distribution that follows from the QCD 
evolution equations. In particular, we neglect the contribution of all
diffraction dissociation processes to the inelastic final state that 
cannot be given as the decomposition of the $\bar q q $ wave function.
For example, we neglect the so called ``fan" diagrams (see Fig.\ref{Fig.3}c) 
which 
give the dominant contributions at very large values of $Q^2$ and small 
$x$ \cite{GLR}.

4. At large value of $b_{\perp}\,>\,b_{0 \, \perp }$,  
$\O$ falls down as $\O\,\propto\,e^{ - 
\,2\,\mu \,b_{\perp}}$  where $\mu $ is the mass of the lightest hadron 
( pion ). Assuming that the DIS cross section cannot increase
 faster than $( \frac{1}{x} )^N$, where power $N = 1$ 
comes from analyticity and crossing \cite{FRST} \cite{MAR},  one can obtain the 
estimate for value of $b_{\perp\, 0}$. Indeed, 
\beq \label{129}
\O\, 
|_{b_{\perp}\,>\,b_{\perp\,0}}\,\,\rightarrow\,\,\frac{1}{x^N}\,e^{- 2 
\mu b_{\perp}}\,\,<\,\,1
\eeq
gives $b_{\perp \,0}\,=\,\,\frac{N}{2\,\mu} \,\ln\frac{1}{x} 
\,+\,O(\frac{1}{\ln(1/x)})$. 

\section{Froissart boundary for $F_2$.}
Actually, \eq{129} gives us the Froissart boundary for $F_2$. 
Differentiating \eq{F2} over $\ln Q^2$ we obtain (for $N_c = 3$ and  $N_f = 3 $)
\beq \label{21}
\frac{ \partial F_2(x,Q^2)}{\partial \ln Q^2}\,\,=\,\,\frac{\, 
Q^2}{3\,\pi^3}
 \int \,d^2 b_{\perp}\,
 \,\{\,1\,\, -\,\, \,e^{- \frac{1}{2} \,\O ( x,\frac{1}{Q^2},b_{\perp} 
)}\,\}\,\,.
\eeq
Using \eq{129} we derive the estimate
\beq \label{22}
\frac{ \partial F_2(x,Q^2)}{\partial \ln Q^2}\,\,<\,\frac{ \,Q^2 
\,b^2_{\perp \,0}}{3 \pi^2}\,\,=\,\,\frac{\,Q^2\,N^2}{12 \pi^2 
\,\mu^2}\,\ln^2\frac{1}{x}\,\,\approx\,\,0.4 \,Q^2 \,\ln^2 
\frac{1}{x}\,\,,
\eeq
where $Q^2$ is in $GeV^2$.

\begin{figure}
\begin{tabular}{cc}
\psfig{file=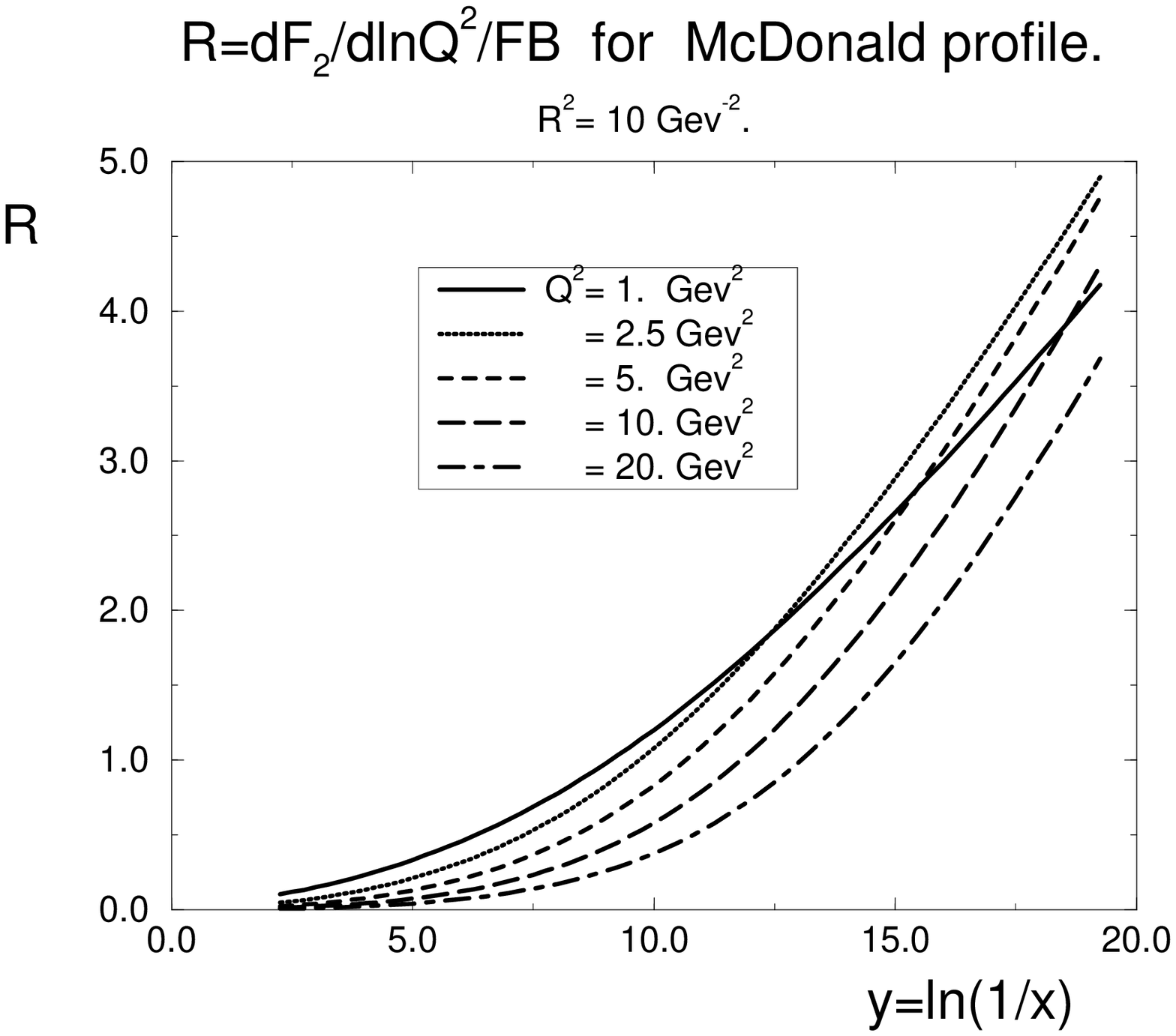,width=70mm} & \psfig{file=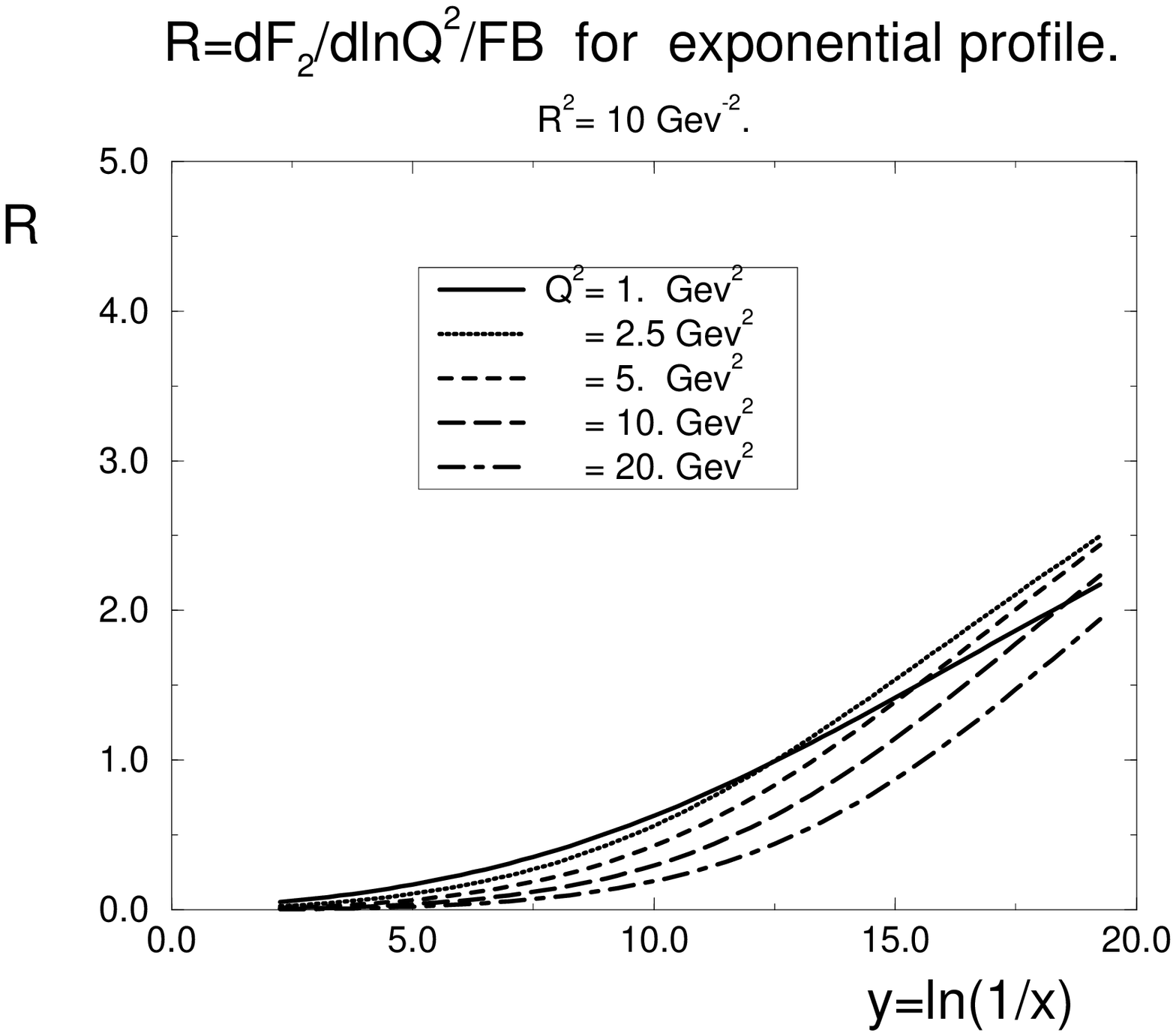,width=70mm}\\
\end{tabular} 
\caption{ The comparison of  $\frac{ \partial F^{GRV}_2(x,Q^2)}{\partial \ln Q^2}$
given by \protect\eq{21} with the Froissart boundary (\protect\eq{25} 
and \protect\eq{26}) with the GRV parametrization.
R = $\frac{\frac{\partial  F_2 (x, Q^2)}{\partial \ln 
Q^2}}{ FB}$, where $FB$ is the Froissart boundary given by 
\protect\eq{25} and \protect \eq{26}.}

\label{Fig.4}
\end{figure}

This boundary turns out to be well above all experimental observations.
 However, we can use a more detailed experimental information to obtain 
more restrictive estimate. Indeed, the HERA data on diffractive photo 
and lepto production of vector mesons \cite{HERAVECTOR} supports the 
idea that $t$-dependence of the DIS amplitude can be factorized out in 
the
form of \eq{125} or, in other words, as $ F_2(x,Q^2;t)\,=\,
F_2 ( x,Q^2) \cdot F (t)$ where
the slope in $t$ corresponds \eq{127} and/or \eq{128}.
Using such  form we can obtain directly from \eq{21} the boundary
\beq \label{23}
\frac{ \partial F_2(x,Q^2; b_{\perp})}{\partial \ln Q^2}\,\,=\,\,\frac{\, 
Q^2}{3\,\pi^3}
 \,\{\,1\,\, -\,\, \,e^{- \frac{1}{2} \,\O ( x,\frac{1}{Q^2},b_{\perp} 
)}\,\}\,\,.
\eeq
Now, using 
$$
F_2(x,Q^2;b_{\perp}) \,\,=\,\,F_2(x,Q^2) \,S( b_{\perp} )\,\,<\,\,F_2(x, 
Q^2 )\,S(0)\,\,,
$$
we derive from \eq{23}
\beq \label{24}
\frac{ \partial F_2(x,Q^2)}{\partial \ln Q^2}\,\, < \,\,\frac{\, 
Q^2}{3\,S(0)\pi^3}\,\,\,, 
\eeq
which gives  for the profile function from \eq{127}
\beq \label{25}
\frac{ \partial F_2(x,Q^2)}{\partial \ln Q^2}\,\, < \,\,\frac{\, 
Q^2}{6\,\pi^2}\,R^2\,\,;
\eeq
while for \eq{128} we have
\beq \label{26}
\frac{ \partial F_2(x,Q^2)}{\partial \ln Q^2}\,\,< \,\,\frac{\, 
Q^2}{3\,\pi^2}\,R^2\,\,.
\eeq
Taking $R^2 \,=\,10 \,GeV^{-2} $ which corresponds both to soft high 
energy phenomenology \cite{GLMSOFT} and the experimental data on 
diffractive  lepto and photo production of vector mesons 
\cite{HERAVECTOR} we are able to compare \eq{25} and \eq{26} with the 
experimental data on $\frac{ \partial F_2(x,Q^2)}{\partial \ln Q^2}$
( see Refs.\cite{EXPSLOPEQ} \cite{DURHAMSLOPEQ} \cite{BH}). In Fig.\ref{Fig.4}
we plot the ratio $R\,=\,\frac{\frac{\partial F_2(x,Q^2)}{\partial \ln 
Q^2}}{FB}$, where $FB$ is the Froissart boundary taking in the form of
 the \eq{25}  or \eq{26}.  For $F_2(x,Q^2)$ we used the GRV 
parametrization which fit the data quite well. One can see that
the GRV parametrization reaches the unitary boundary   ( $R\,=\,1 $ )
at $Q^2= Q^2_0 = 2 - 4 
\,GeV^2$ at HERA kinematic region.
 We can  estimate the value of $Q^2_0$ even  more accurately using 
the
parameterization of the experimental data on $F_2$ given in 
Ref.\cite{BH}, namely, $\frac{\partial F_2(x, Q^2)}{\partial \ln 
Q^2}\,=\,0.364\,\log\frac{0.074}{x}$. Comparing this parameterization 
with \eq{26} one obtains $Q^2_0\,=\,1.092 
\frac{\pi^2}{R^2}\,\log\frac{0.074}{x}$. Therefore, the value of $Q^2_0$ 
turns out to be pretty high at low $x$.
 This fact encourage us to search for  a  more 
microscopic approach for the parton - parton interaction in the parton 
cascade at moderate values of $Q^2\,\approx\,2\,GeV^2$.

\section{ Froissart boundary for the gluon structure function.}
As has been pointed out by A. Mueller \cite{MU90}, the gluon structure 
function can be also written through the dipole $G G$ - pair interaction 
with a target in a similar way as has been done for $ F_2$.
Omitting all calculation that can be found in Refs. \cite{MU90} 
\cite{AGL}, one can derive
\beq \label{41}
\frac{\partial^2 xG( x,Q^2 )}{\partial \ln Q^2\,\partial \ln 
\frac{1}{x}}\,\,=\,\,\frac{2\,Q^2}{\pi^3} \,\int d^2 b_{\perp}\,
\{\,1\,\,-\,\,e^{ - \frac{1}{2}\,\O_{GG} ( x, \frac{1}{Q^2}, b_{\perp} 
)}\,\}\,\,,
\eeq
where the opacity $\O_{GG}$ for $GG$ - dipole scattering has the same 
properties ( see section 2.2) as for $\bar q q $- dipole scattering. 
The difference is that in the limit of small $\O_{GG}$,  $ \O_{GG}\,=\,
\frac{9}{4} \,\O_{\bar q q}$ for $N_c = 3 $.  Repeating all arguments of 
section 3 one can obtain the Froissart boundary for $ x G ( x,Q^2 )$ in 
the form
\beq \label{42}
\frac{\partial^2 xG( x,Q^2 )}{\partial \ln Q^2\,\partial \ln 
\frac{1}{x}}\,\,<\,\,\frac{2\,Q^2}{\pi^2} \, b^2_{\perp\,0}\,
\,=\,\,\frac{2 \,Q^2}{4\,\mu^2\,\pi^2}\,\ln^2 \frac{1}{x}\,\,=\,\,
2.5 \,Q^2\,\ln^2\frac{1}{x}\,\,.
\eeq 
\begin{figure}
\begin{tabular}{cc}
\psfig{file=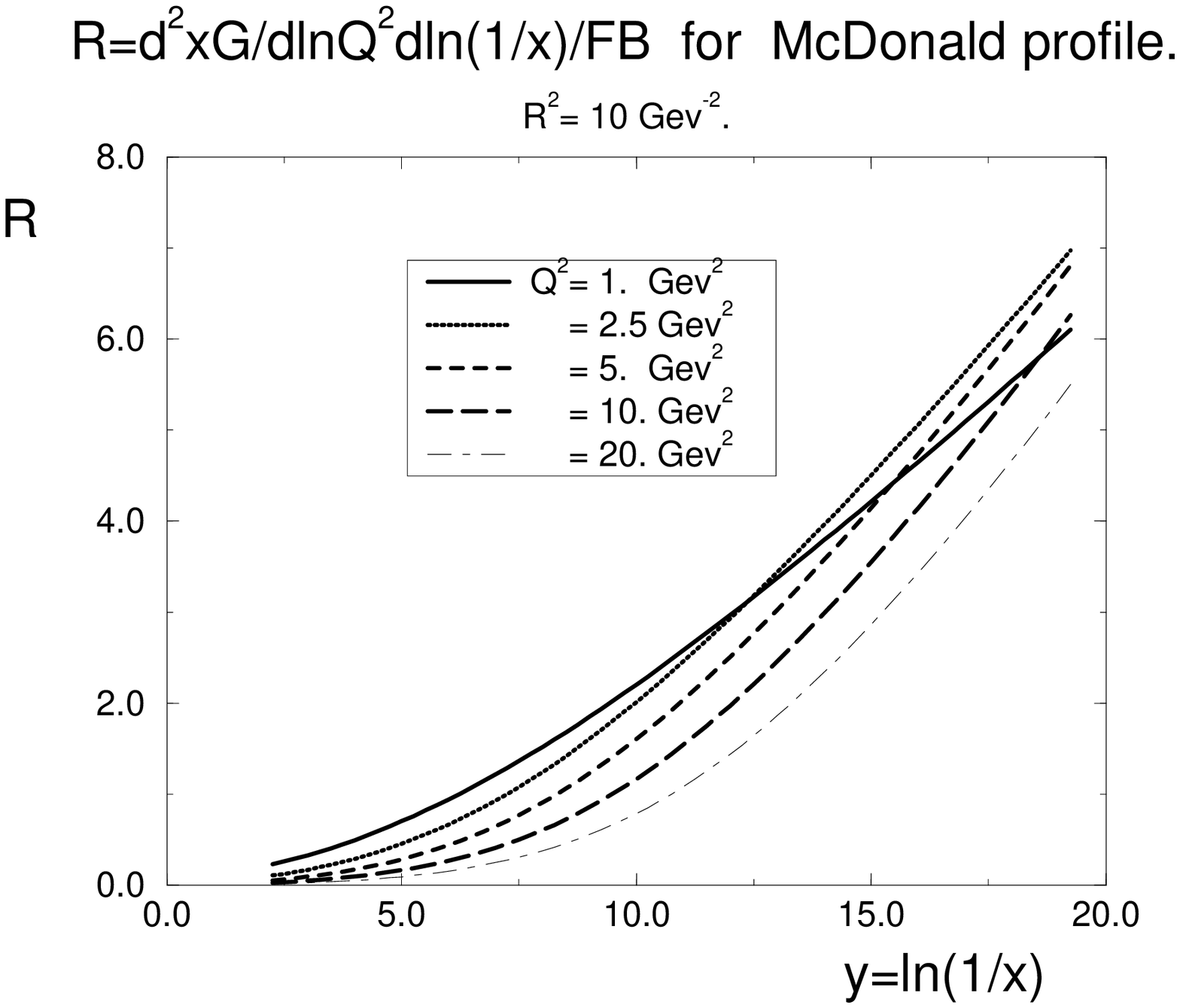,width=70mm} & \psfig{file=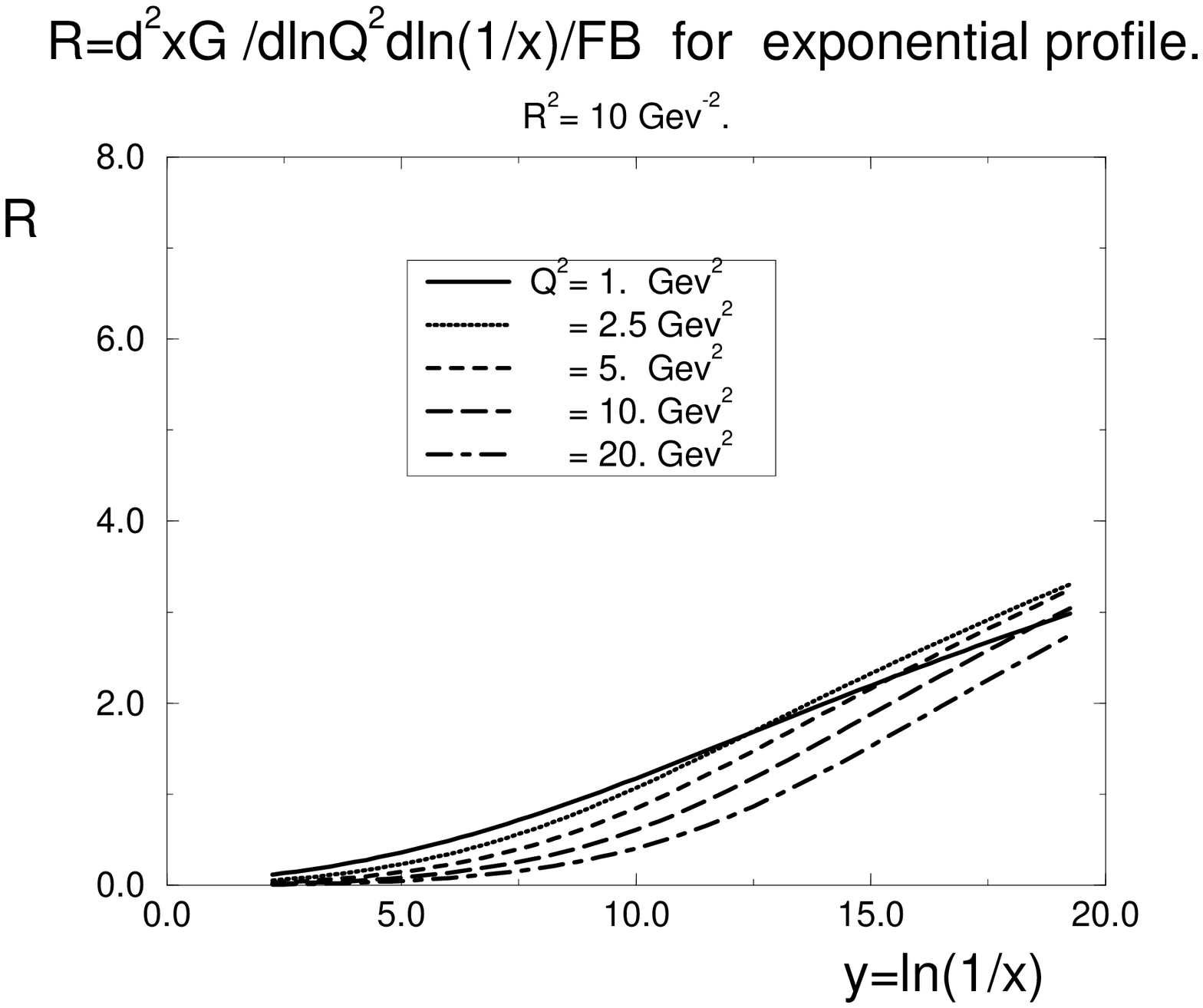,width=70mm}\\
\end{tabular} 
\caption{The comparison of 
 $\frac{\partial^2 xG^{GRV}( x,Q^2 )}{\partial \ln Q^2\,\partial \ln 
\frac{1}{x}}$ given by \protect\eq{41} 
with the Froissart boundary given by \protect\eq{43} and \protect\eq{44}.
R = $\frac{\frac{\partial^2 x G (x, Q^2)}{\partial \ln 
Q^2 \,\partial \ln \frac{1}{x}}}{ FB}$,
 where $FB$ is the Froissart boundary given by 
\protect\eq{43} and \protect \eq{44}}
\label{Fig.5}
\end{figure}

This boundary is much higher than the
 $\frac{\partial^2 xG( x,Q^2 )}{\partial \ln Q^2\,\partial \ln 
\frac{1}{x}}$ in all current parameterizations of the gluon structure 
function \cite{MRS} \cite{GRV} \cite{CTEQ}. However, using the approach
 developed in the previous section one can obtain more restrictive
estimates, namely
\beq \label{43}
\frac{\partial^2 xG( x,Q^2 )}{\partial \ln Q^2\,\partial \ln 
\frac{1}{x}}\,\,<\,\,\frac{Q^2}{\pi^2}\,R^2
\eeq
 and
\beq \label{44}
\frac{\partial^2 xG( x,Q^2 )}{\partial \ln Q^2\,\partial \ln 
\frac{1}{x}}\,\,<\,\,\frac{2\,Q^2}{\pi^2}\,R^2
\eeq
for $S(b_{\perp})$ from \eq{127} and \eq{128}, respectively.
In Fig.\ref{Fig.5} we plot
 $ \frac{\partial^2 xG( x,Q^2 )}{\partial \ln Q^2\,\partial \ln 
\frac{1}{x}}$ for  the GRV parameterization of the gluon structure 
function and compare them with  \eq{43} and \eq{44}. One can see, that
 the gluon structure function reaches the unitarity limit ( $R$ = 1 ) at HERA 
kinematic region.

 In Ref.\cite{BH} it has been shown that $\frac{\partial 
xG(x,Q^2)}{\partial \ln \frac{1}{x}}\,\approx\,3$ from the experimental 
data on $F_2$. It means that the gluon structure function reaches the 
unitarity boundary at $Q^2_0\,\approx\,\frac{3 \pi^2}{R^2}\,\approx\,3 
GeV^2$ ( see \eq{43} ).

\section{ Parameter for the SC.}

To find out the parameter for the SC let us rewrite \eq{41} in the 
kinematic region where $\O_{GG}\,<\,1$. One obtains 
\beq \label{51}
\frac{\partial^2 xG( x,Q^2 )}{\partial \ln Q^2\,\partial \ln 
\frac{1}{x}}\,\,=\,\,\frac{2\,Q^2}{\pi^3} \,\int d^2 b_{\perp}\,
\{\,\frac{1}{2}\,\O_{GG} ( x, \frac{1}{Q^2}, b_{\perp} 
) \,\,-\,\,\frac{1}{8} \,\O^2_{GG} ( x, \frac{1}{Q^2}, b_{\perp} 
)\}\,\,.
\eeq
Substituting $\O_{GG} \,=\,\frac{N_c \as \pi^2}{Q^2}\,  x G(x, 
Q^2)\,S( b_{\perp})$ we have
\beq \label{52}
\frac{\partial^2 xG( x,Q^2 )}{\partial \ln Q^2\,\partial \ln 
\frac{1}{x}}\,\,=\,\,\frac{ \as\,N_c}{\pi} x G(x,Q^2 )\,\,-\,\,
\frac{\as^2\, N_c^2 \pi^2 }{4 Q^2}  ( x G( x,Q^2 ) )^2  \int d b^2_{\perp} 
\,S^2(b_{\perp})\,\}\,\,,
\eeq
which can be rewritten in the form
\beq \label{53}
\frac{\partial^2 xG( x,Q^2 )}{\partial \ln Q^2\,\partial \ln 
\frac{1}{x}}\,\,=\,\,\frac{N_c \,\as}{\pi} x G( x,Q^2 )\,\cdot\,\{
\,1\,\,-\,\,\frac{\kappa}{4}\,\}\,\,,
\eeq
with 
\beq \label{54}
\kappa\,\,=\,\,\frac{\as \,N_c \,\pi^3}{Q^2}  xG( x,Q^2 ) \int d b^2_{\perp} 
\,S^2(b_{\perp})\,\,.
\eeq
\begin{figure}
\centerline{\psfig{file=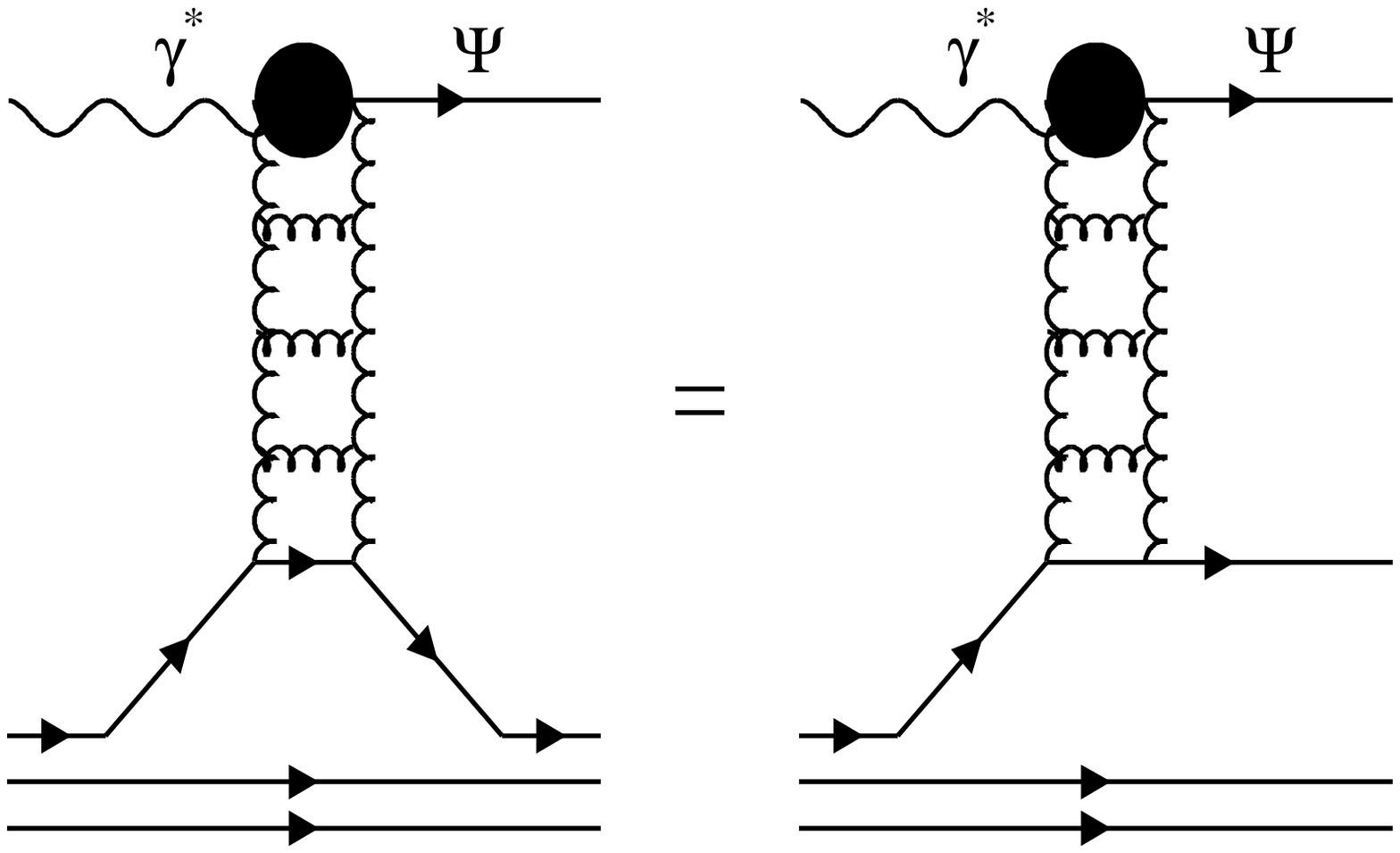,width=90mm}} 
\caption{The $J/\Psi$ production without a) and with b) 
dissociation of the proton.}
\label{Fig.6}
\end{figure}

The above equation gives the same definition for $\kappa$ as \eq{I1} for
exponential form of $S(b_{\perp}$) ( see \eq{128} ).  Using the new HERA 
data on photoproduction of J/$\Psi$ meson
\cite{HERAPSI}  we are able to estimate the 
value of $R^2$ in the definition of $\kappa$ , recalling that $R^2$ is 
the size of the target only in the oversimplified eikonal ( Glauber ) 
model. To illustrate the point we picture in Fig.\ref{Fig.6} the process of 
J/$\Psi$ photoproduction in the additive quark model (AQM ). We see that 
we have two processes with different slopes ($ B$ )  in $t$
 ( or in $b^2_{\perp}$ 
): the J/$\Psi$ production without ( Fig.\ref{Fig.6}a ) ($B_{el}\,=\,
5\, GeV^{-2}$)  and with ( Fig.\ref{Fig.6}b ) 
( $B_{in}\,=\,1.66 \,GeV^{-2}$ ) dissociation of the  proton.
 The AQM gives us the simplest estimates for the resulting slope ( 
$R^2$ ) in \eq{I1} if we neglect any slope from the Pomeron - J/$\Psi$ 
vertex in Fig.\ref{Fig.6}, namely
\beq \label{55}
\frac{1}{R^2}\,\,=\,\,\frac{1}{4}\,\{\,\frac{3}{2 B_{el}} 
\,\,+\,\,\frac{1}{2 B_{in}}\,\}\,\,\approx\,\,\frac{1}{5} \,GeV^{-2}\,\,.
\eeq
This is a reason why we used $R^2 = 5  \,GeV^{-2}$ in  Fig.\ref{Fig.1} 
to estimate 
the scale for the SC.

In our estimates of the value of the deep inelastic structure functions
at $b_{\perp}$ = 0 ( see Eq.(27) ) we used an assumption that the SC 
does not change the value of R. To justify this assumption we plot in 
Fig.7 the $x$-dependence of the average $b^2_{\perp}$ calculated in the 
Glauber (eikonal ) approach with $R^2 = 5\, GeV^{-2}$. One can see that
 $< b^2_{\perp} > $ only weakly depends on $x$ in the HERA kinematic 
region.
\begin{figure}
\centerline{\epsfig{file=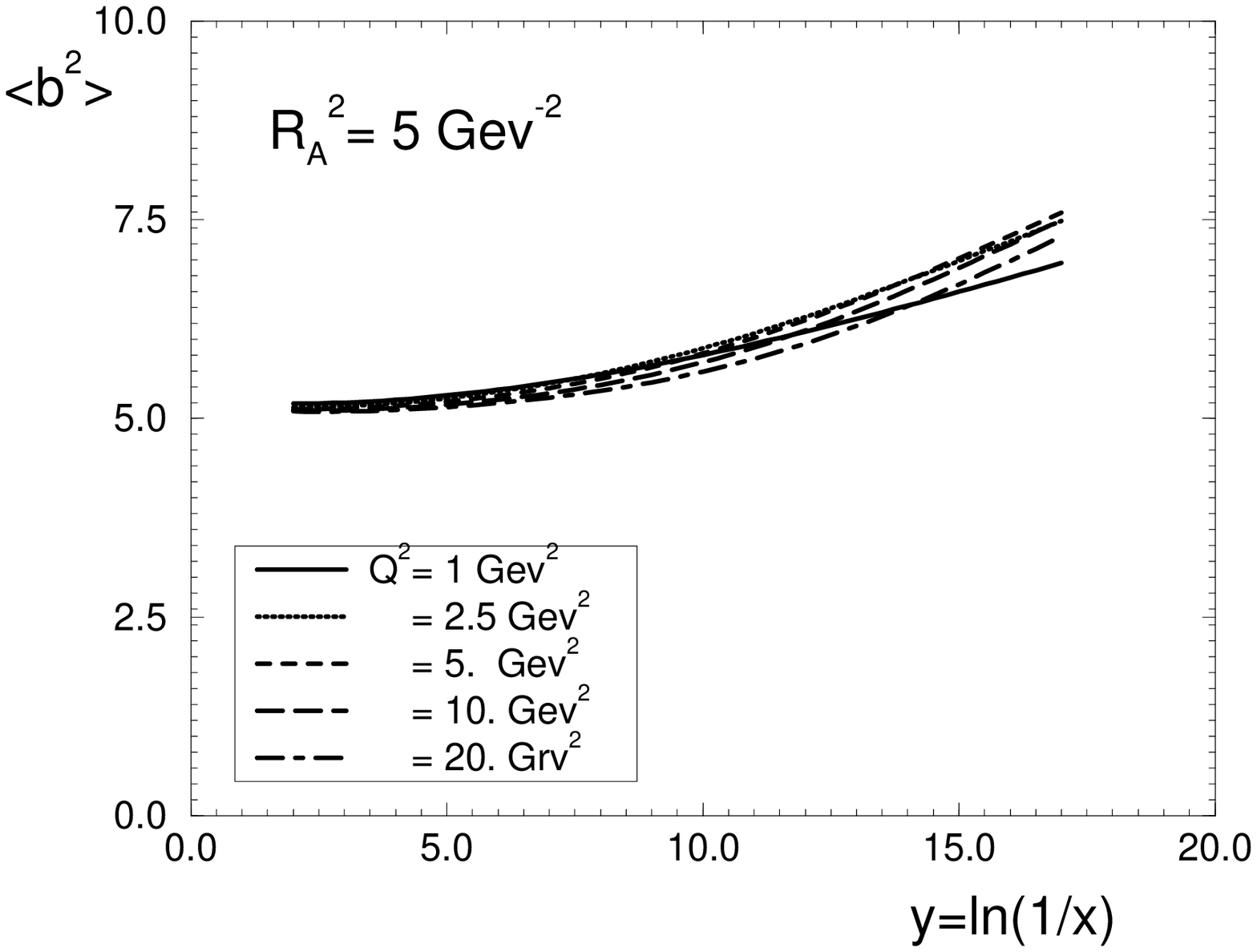,width=90mm}} 
\caption{The $x$ - dependence of $< b^2_{\perp}> $ in the Glauber 
(eikonal) approach with $R^2\,=\,5 GeV^{-2}$.}
\label{Fig.7}
\end{figure}

\section{Summary.}
It has been presented the derivation
 of the Froissart boundary for the deep inelastic 
structure functions.

The comparison of the Froissart boundary with HERA experimental data
shows that both $F_2(x,Q^2)$ and $x G(x,Q^2)$ hit
the unitarity limit at $Q^2\,\approx\, 2 - 4 GeV^2 $. This fact gives 
rise to a challenge for theoreticians to explain the matching between 
the deep inelastic scattering and real photoproduction process in the 
framework of QCD. 

We hope that this letter as well as Ref.\cite{BH} will stimulate 
the new round of the discussions on the theory of the shadowing 
corrections in the deep inelastic processes. We believe that the 
resolution of all difficulties could be found assuming that the SC has 
worked in the full in the gluon structure function and has been taken
in the phenomenological initial gluon distribution in standard 
parameterizations \cite{GRV} \cite{MRS} \cite{CTEQ}. However, much more 
work is needed to prove this.
 
This work was supported in part by the U.S. Department of Energy, 
Division of High Energy Physics, Contract No. W-31-109-ENG-38 and by 
CNPq,CAPES and FINEP,Brazil.

\end{document}